\documentclass{article}
\usepackage[dvips]{graphicx}
\begin{document} 
%
\begin{center}
{\large   }
\end{center}
\vspace{2 ex}
 
  
\begin{center}

\LARGE\bf
\rule{0mm}{7mm} Leading Effects in Hadroproductions of
$\Lambda_c$ and $D$ From Constituent Quark-Diquark Cascade Picture\\
\end{center}

\vspace{4ex}
\begin{center}
Tsutomu Tashiro$^1$, Shin-ichi Nakariki,$^1$, Hujio Noda$^2$, Kisei Kinoshita$^3$ and
Shuxin Lan$^4$ \\
\vspace{3ex}
$^1$Department Simulation Physics, Okayama University of Science, Ridai-cho, Okayama 700-0005, Japan \\
$^2$Department of Mathematical Science, Ibaraki University, Bunkyou, Mito 310-0056, Japan  \\
$^3$Physics Department, Kagoshima University, Korimoto, Kagoshima 890-0065, Japan\\
$^4$Department Physics, Okayama University, Okayama 700-0005, Japan \\
\end{center}
\vspace{2 ex}

\centerline{}

\vspace{1 ex}

\begin{abstract}
We discuss the hadroproductions of $\Lambda_c, \bar{\Lambda}_c, D$ and $\bar{D}$ in the framework of 
the constituent quark-diquark cascade model taking into account the valence quark annihilation. 
 The spectra of $\Lambda_c$ and $\bar{\Lambda}_c$ in $pA, \Sigma^-A$ and 
$\pi^-A$ collisions are well explained by the model using the values of parameters 
used in hadroproductions of $D$ and $\bar{D}$. 
It is shown that the role of valence diquark in the incident baryon is 
important for $\bar{D}$ productions as well as for $\Lambda_c$ production. 

\end{abstract}



\section{Introduction}

Hadroproductions of charmed particles have been measured in fixed-target experiments
\cite{exp,e791}  
and indicate the large leading/non-leading asymmetries, 
i.e. asymmetries between production
cross sections of leading particles and those of non-leading particles defined as
\begin{eqnarray}
A=\frac{\sigma(leading)-\sigma(non-leading)}{\sigma(leading)+
\sigma(non-leading)}.
\end{eqnarray}
 The leading particle shares a valence quark in 
common with the incident hadron, while the non-leading one does not. 
The leading particles are copiously produced at large Feynman $x$ as compared with 
the non-leading particles in the forward region of the incident hadron. This is called as 
leading particle effect.  For example, 
the asymmetry between $D^-(d\bar{c})$ and $D^+(c\bar{d})$ 
in $\pi^-(d\bar{u}) $ interaction with a nucleon is defined as
$
A_{\pi^- N}(D^-,D^+) = (d\sigma(D^-)-d\sigma(D^+))/(d\sigma(D^-)+d\sigma(D^+))
$
and increases from zero to nearly one with $x$ in the $\pi^-$ fragmentation region.  
In the conventional perturbative QCD at leading order (LO), the factorization theorem predicts that 
$ c $ and $ \bar{c} $ quarks are produced symmetrically and 
then fragment into $D$ and $\bar{D}$ independently.   As a consequence, LO 
perturbative QCD predicts 
no asymmetry in contrast with the experimental data. The next to leading order (NLO) calculations  
generates asymmetries but they are much smaller than the data\cite{nde,fmnr}. 

To explain the leading/non-leading asymmetry, many attempts have 
been investigated: 
$k_T$ factorization\cite{rss}, 
string fragmentation\cite{PYTHIA,piskounova,arakelyan_volko}, 
intrinsic charm 
contributions\cite{intrinsic}, 
recombination process with beam remnants\cite{bednyakov,c_h_magnin,LS,LS02,shabelski,A_Magnin_M-N}, 
recombination with participants in the hard scattering process\cite{bjm_hdrn,bkjm}, 
recombination with surrounding light quarks in projectile and target\cite{Rapp_Shuryak}, 
recombination using valon concept\cite{hwa}, 
light quark fragmentation\cite{DD}, meson cloud model\cite{cdnn} 
and so on. 
Productions of $\Lambda_c$ and $\bar{\Lambda}_c$ in $\Sigma^-A$ 
\cite{e791,wa89} and $\Sigma^-A, pA$ and $\pi^-A$ collisions\cite{SELEX} have been  
measured and analyzed in Refs.\cite{LS02,A_Magnin_M-N,bkjm,a_h_magnin_s} and \cite{piskounova_Lambda_c}. 
These models explain the leading/non-leading asymmetry of charmed hadrons 
successfully, however, there are some uncertainties in the 
perturbative QCD calculations as reviewed 
in Refs.\cite{fmnr97} and \cite{ac}. 
Hadronization of heavy quarks is still an open problem, since 
it is intrinsically non-perturbative process. 

We have applied the covariant quark-diquark cascade model for hadroproductions of 
charmed mesons as well as hadrons composed $u, d$ and $s$ quarks\cite{qdq,qdqA,tnnik,tnkn}.
In the present paper,
we investigate the hadroproductions of $\Lambda_c, \bar{\Lambda}_c, D$ and $\bar{D}$ 
and the leading/non-leading asymmetry in the framework of 
the constituent quark-diquark cascade picture and discuss the role of constituent diquark. \\

\section{The model}

\subsection{Valence quark distributions in the incident hadron}
\label{sec:quark_dstb}

\begin{figure}[h]
\center
\includegraphics[height=4cm]{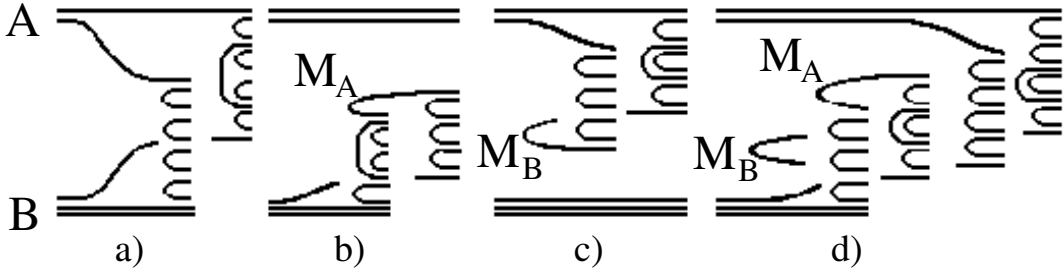}
\vspace*{8pt}
\caption{The interaction mechanism in $AB$ collision: (a) Non-diffractive 
dissociation type, (b), (c) Single-diffractive and (d) double diffractive 
dissociation type mechanisms.}
\label{fgr:intrctn_type}       
\end{figure}

We assume that a baryon is composed of a constituent quark and a constituent diquark, and 
a meson is composed of a constituent quark and a constituent anti-quark.  The constituent quark (diquark) 
is a quark-gluon cluster consisting of the valence quark (diquark), sea quarks and gluons.  
Hereafter, we simply call them as quark and diquark.  
When the collision between the incident hadrons $A$ and $B$ occurs, it is assumed that the 
incident hadrons either break up into two constituents or 
emit gluons followed by a quark-antiquark pair creation.  
There are four interaction types: a) non-diffractive dissociation, b) and c) 
single-diffractive dissociations of $ B$ and $A$, and d) double-diffractive dissociation 
types as shown in Fig.\ref{fgr:intrctn_type}.  The probabilities of these types to occur 
are $ (1 - P_{gl})^2 , P_{gl} (1 - P_{gl}), P_{gl} (1 - P_{gl})$ and $ P_{gl}^2$, 
respectively.  Here we denote the quark-antiquark pair emitted from $ A $ ($ B $) 
via gluons as $ M_A $ ( $ M_B $).  The probabilities of $ M_A $ ( $ M_B $) 
to be $u\bar{u}, d\bar{d}, s\bar{s} $ and $ c\bar{c} $  are 
are denoted by 
$  P_{u\bar{u}}, P_{d\bar{d}}, P_{s\bar{s}}$ and 
$ P_{c\bar{c}}$, respectively.  

The light-like variables of $ A $ and $ B $ are defined 
as follows:
\begin{eqnarray}
x_{0\pm}^A=\frac{E^A \pm p_{cm}}{\sqrt{s_0}},~~~x_{0\pm}^B=\frac{E^B \mp p_{cm}}{\sqrt{s_0}} ,
\end{eqnarray}
where $ \sqrt{s_0}=E^A+E^B $ is the center of mass energy of the incident hadrons $A$ and 
$B$.
The momentum fractions of $A,B$ and $M_A$ for the process shown in Fig.1b 
are given as follows: The momentum fraction of $M_A$ is fixed by the distribution function 
\begin{eqnarray}
  H_{M_A/A}(z)  = z^{\beta_{gl}-1}(1-z)^{\beta_{ld}-1}/B(\beta_{gl},\beta_{ld}) ,
\label{eqn:Hmaa}
\end{eqnarray}
and the uniform distribution $R$ in the interval from zero to one as,
\begin{eqnarray}
x_+^{M_A}=x_{0+}^Az,~~~x_-^{M_A}=x_{0-}^AR.
\label{eqn:Hmaa_2}
\end{eqnarray}
Then the momentum fractions of the incident particles $A$ and $B$ become as follows: 
\begin{eqnarray}
x_+^{A}=x_{0+}^A(1-z),~~x_-^{A}=m_A^2/(x_+^{A}~s_0),
\nonumber
\end{eqnarray}
\begin{eqnarray}
x_-^B=x_{0-}^B-(x_{-}^A-x_{0-}^A(1-R)),~~x_+^B=x_{0+}^B,
\label{eqn:Hmaa_2_leading}
\end{eqnarray}
where the mass shell condition is considered and transverse momenta are neglected. 
Analogous expressions describe the momentum fractions of $A,B$ and $M_B$ for the process shown in Fig.1c
, upon replacing superscripts 
$A \leftrightarrow B$ and subscripts
$+ \leftrightarrow -$.
The single-diffractive dissociation type mechanisms result in  diffractive peaks of spectra of the 
same kind of the incident particles $A$ and $B$ at $|x| \approx 1$. 
 The probability of the 
incident hadron to emit gluons followed by a quark-antiquark pair 
$ P_{gl}$ and the parameters $\beta_{gl}$ and $\beta_{ld}$ in (\ref{eqn:Hmaa})
are chosen to reproduce the diffractive peaks of the spectra such as $\pi^\pm, K^\pm$ and $p$ in 
$\pi^\pm p,~K^\pm p$ and $pp$ collisions at $|x| \approx 1$\cite{tnnik}.  
The parameters are chosen as $ P_{gl}=0.15, \beta_{gl}=0.1$ and 
$\beta_{ld}=3.0$. 
However, the main contributions to charmed hadrons come from the non-diffractive dissociation type mechanism.

The distribution functions of the constituents in the projectile $ A $ 
composed of $ a $ and $ a' $ are described as
\begin{eqnarray}
 H_{a/A}(z) = H_{a'/A}(1-z) = \frac{z^{\beta_a-1}(1-z)^{\beta_{a'}-1}}{B(\beta_a,\beta_{a'})}. 
\label{eqn:Ha/A}
\end{eqnarray}
Similarly, the distribution function of $q$ in $M_A$ is given as
$ H_{q/M_A}(z) = z^{\beta_q-1}(1-z)^{\beta_q-1}/B(\beta_q,\beta_q)$. 
The dynamical parameters $ \beta$'s in (\ref{eqn:Ha/A}), which 
determine the momentum sharing of the constituents, are 
related to the intercepts of the Regge trajectories as 
$\beta_u=\beta_d=1-\alpha_{\rho-\omega}(0)\approx 0.5,~~ \beta_s=1-\alpha_\phi(0)\approx1.0,
\beta_c=1-\alpha_{J/\psi}(0)\approx4.0$\cite{mnkt,cthkp}.  For flavor anti-symmetric diquark $[qq']$ 
and symmetric diquark $\{qq'\}$, the dynamical parameters $\beta_{[qq']}$ and 
$\beta_{\{qq'\}}$ are chosen as $\beta_{[qq']}=\gamma_{[\ ]}(\beta_q+\beta_{q'})$ and 
$\beta_{\{qq'\}}=\gamma_{\{\}}(\beta_q+\beta_{q'})$, respectively. 
 The commonly shared constituents of $\Lambda_c$ with the incident $p, \Sigma^-$ and $\pi^-$ 
beams are $ud$-diquark, $d$ and $d$-quark respectively. 
$\bar{\Lambda}_c$ has a commonly shared $\bar{u}$ quark with $\pi^-$ and $K^-$ beams and 
$\Lambda_c$ is a non-leading particle in $K^-$ beam. 
These distributions of the constituents in the incident hadrons affect the shapes of spectra 
of produced hadrons significantly. 
Hadrons are produced on the chain between a constituent valence quark($q_A$)(anti-quark($\bar{q}_A$))
from the incident hadron $A(M_A)$ and the constituent valence diquark($qq_B(\bar{q}_B)$)(quark($q_B$)) from the 
incident hadron $B(M_B)$ for a meson-baryon collision as illustrated in Fig.1.  

The small intrinsic transverse momenta 
of constituents in the incident hadrons are introduced by changing the direction of 
the incident constituents 
by an angle $\varphi_{in}$ in the rest frame of the chain.  Here we choose the 
distributions for $z=\cos\varphi_{in}$ as 
\begin{eqnarray}
D_{in}(z)=\frac{\beta_{in}+1}{2^{\beta_{in}+1}}(1+z)^{\beta_{in}} 
\label{eqn:intrinsic} 
\end{eqnarray}
in the region $-1 < z < 1$. 
Here we choose very small intrinsic transverse momenta and use the value $\beta_{in}=40$ used in Ref.\cite{tnkn}  
in order to reproduce $p_T^2$ dependences of $D$ mesons.

We take account of the annihilation of the valence antiquark from beam and valence quark 
from target in a slightly different way from the one pointed out in Refs.
\cite{bednyakov} and \cite{c_h_magnin}.  
When the incident constituents in $A$($M_A$) and $B(M_B)$ 
are quark and 
its anti-quark, it is assumed that the annihilation 
process occurs with 
the probability $P_{an}$ and the non-annihilation process occurs with 
$1-P_{an}$. 
Let us consider the case of $\pi^-p$ collision.
The annihilation process
\begin{eqnarray}
\bar{u} + u \rightarrow \bar{q} + q \rightarrow "hadrons" + "hadrons"
\label{eqn:anni} 
\end{eqnarray}
occurs with the probability $P_{an}$, when $\bar{u}$ in $\pi^-$ interacts 
with $u$ in $p$.
Branching ratios of $\bar{u}u \rightarrow 
\bar{u}u, \bar{d}d, \bar{s}s,$ and $\bar{c}c$ are chosen to be equal to $P_{u\bar{u}}, 
P_{d\bar{d}}, P_{s\bar{s}}$ 
and $P_{c\bar{c}}$ for the channels allowed energetically. 
The produced $\bar{q}$ and $q$ in (\ref{eqn:anni}) 
are supposed to be non-free in terms of hadronization mechanism.  Thus it is assumed 
that $\bar{q}$ has tendency to be produced in the forward direction of $\bar{u}$ in 
$\pi^-$ beam. Therefore, 
 here we choose the distributions for $z=\cos\varphi_{an}$ as 
\begin{eqnarray}
D_{an}(z)=\frac{\beta_{an}+1}{2^{\beta_{an}+1}}(1+z)^{\beta_{an}} 
\label{eqn:annidrctn} 
\end{eqnarray}
in the region $-1 < z < 1$, 
where
$\varphi_{an}$ is the angle between the directions of $\bar{u}$ and $\bar{q}$.
To reproduce the $A_{\pi^- N}(D^0,\bar{D}^0)$, 
we choose the value $P_{an}=0.2$ and use a value of $\beta_{an}=20$ 
which roughly means $D_{an}(z) \sim \delta(z-1)$
\cite{tnkn}.  \\

\subsection{Quark-diquark cascade process}

Hadrons are produced on the cascade chain by processes
\noindent
\begin{eqnarray}
\label{eqn:cscdq2M}
q & \rightarrow & M(q \bar{q}')+q',\\
\label{eqn:cscdq2B}
q & \rightarrow & B(q[q'q''])+\overline{[q'q'']} ,  B(q\{q'q''\})+\overline{\{q'q''\}},\\
\label{eqn:cscda2B}
\overline{[q'q'']} & \rightarrow & \overline{B}(\bar{q}\overline{[q'q'']})+q,\\
\label{eqn:cscda2M}
\overline{[q'q'']} & \rightarrow & M(q\bar{q}')+\overline{[qq'']} , M(q\bar{q}')+\overline{\{qq''\}},\\ 
\label{eqn:cscds2B}
\overline{\{q'q''\}} & \rightarrow & \overline{B}(\bar{q}\overline{\{q'q''\}})+q,\\
\label{eqn:cscds2M}
\overline{\{q'q''\}} & \rightarrow & M(q\bar{q}')+\overline{[qq'']} ,M(q\bar{q}')+\overline{\{qq''\}},
\end{eqnarray}
where $q$ denotes $u, d, s $ and $c$ and $[q'q'']$ does $[ud], [us],[uc],[ds],[dc], 
[sc]$ and so on.  The symbols $[\ ]$ and $\{\ \}$ denote the flavor anti-symmetric and 
symmetric diquarks, respectively. 
 Meson production probabilities 
from $q, \overline{[q'q'']} $ and $ \overline{\{q'q''\}}$ are $ 1-\epsilon, \eta_{[\ ]} $ 
and $ \eta_{\{\ \}} $, respectively. The parameter $\epsilon$ is the probability of baryon 
productions from a quark and is related to the baryon productions from $\pi$ and $K$ 
beams resulting in $\epsilon=0.07$.\cite{qdq} 
The processes (\ref{eqn:cscda2M}) and (\ref{eqn:cscds2M}) contribute to production of 
baryons which share no valence quark or diquark in common with the incident hadron. 
Although both $\Omega$ and $\bar{\Omega}$ share no valence quark common with the incident proton, 
the spectrum of $\Omega$ is harder than that of $\bar{\Omega}$ in the proton fragmentation region.
In our model this is described as follows: after emitting $K$ mesons through these processes, 
the incident diquark in the incident proton converts into $\{ss\}$ diquark and produce $\Omega$ by the process 
(\ref{eqn:cscds2B}).
The difference between the spectra of $\Omega$ 
and $\bar{\Omega}$ in $pp$ collision is described by the processes  (\ref{eqn:cscda2M}),(\ref{eqn:cscds2M}) 
and (\ref{eqn:cscds2B}) and the parameters are chosen as $\eta_{[]}=\eta_{\{\}}=0.25$.\cite{qdq,tnkn}

For the cascade process $ a \rightarrow H(a\bar{b}) + b$ 
in the beam (target) side, the distribution function of light-like fraction of $b$ 
is assumed as
\begin{eqnarray}
 F_{ba}(z) = z^{\gamma\beta_a-1}(1-z)^{\beta_a+\beta_b-1} /B(\gamma\beta_a,\beta_a+\beta_b) 
\label{eqn:Fqq} 
\end{eqnarray}
and the light-like fraction of $b$ becomes $x_+^b=zx_+^a ~(x_-^b=zx_-^a)$.
We fix the transverse momentum of the hadron $H$ as a difference between two random points 
${\mbox{\boldmath$p$}_{T}^H}=\mbox{\boldmath$p$}_{T2}-\mbox{\boldmath$p$}_{T1}$ with their
lengths fixed by the distribution function 
\begin{eqnarray}  
G(\mbox{\boldmath$p$}_{T}^2)=\frac{\sqrt{m_H}}{C}\exp(-\frac{C}{\sqrt{m_H}}\mbox{\boldmath$p$}_{T}^2) 
\label{eqn:pT2} 
\end{eqnarray}
in ${\mbox{\boldmath$p$}_{T}^2}$ space, where $m_H$ denotes the mass of $H$.  
The parameter 
$C$ is fixed from the experimental data on ${\mbox{\boldmath$p$}_{T}^2}$ distributions of pions\cite{tnnik}.  
We use the values of $\gamma=1.75$ and $C=1.8$GeV$^{-\frac{3}{2}}$, from previous analyses\cite{tnnik,tnkn}. 

  We now see how the cascade processes occurs between the incident constituents, for example, 
$q$ from $A$ $(M_A)$ and $\{q'q''\}$ from $B$.
We redefine the light-like fractions of $q$ and $\{q'q''\}$ in the rest frame of the incident constituents 
$q$ and $\{q'q''\}$.
The momentum sharing of the cascade process $ q + \{q'q''\} \rightarrow H(q\bar{b}) + b + \{q'q''\}$
from $q$ with $x_\pm^q$ and $\{q'q''\}$ with $x_\pm^{\{q'q''\}}$ takes place as follows\cite{qdqA,km}: 
Using Eq.(\ref{eqn:Fqq}) with $a=q$, we fix the light-like fractions of $b$ and $H(q\bar{b})$ as 
$x_+^b=x_+^q z$ and $x_+^H= x_+^q-x_+^b$, respectively and put $x_-^b=x_-^q$. 
The transverse momentum of $H$ is fixed by Eq.(\ref{eqn:pT2}). Then $x_-^H$ is fixed from the 
onshell condition $x_+^H x_-^H =(m_H^2 + {\mbox{\boldmath$p$}_{T}^2})/s'$, where $\sqrt{s'}$ is 
the subenergy of the incident $q$ and $\{q'q''\}$ system. The transverse momentum of $ b $ 
is $ {\mbox{\boldmath$p$}}^{b}_T = {\mbox{\boldmath$p$}}^q_T - {\mbox{\boldmath$p$}}_{T}^H $.  
The light-like fraction of $ \{q'q''\} $ is decreased to 
$ \tilde{x}^{\{q'q''\}}_{-} = x^{\{q'q''\}}_{-} - x^{H}_{-}$.  If the energy of the $b + \{q'q''\}$ system is 
enough to create another hadron, the cascade process such as 
$ b + \{q'q''\} \rightarrow b + H(c\{q'q''\}) + \bar{c} $ takes place in the opposite side.
In the final step, we assume that the constituents recombine into one or two hadrons 
according to the processes: $
     q + \bar{q}' \rightarrow M(q\bar{q}'),~ 
     q + [q'q''] \rightarrow B(q[q'q'']), ~
     \{qq'\} + \overline{[q''q''']} \rightarrow M(q \overline{q''}) + M(q'\overline{q'''})$~ 
and so on. The momenta of the recombined hadrons are the sum of those of the final 
constituents and are offshell.  In the cms of incident hadrons $A$ and $B$, we have energy-momentum conservation relations 
$\sum\limits_i {\mbox{\boldmath$p$}}_i=0$ and $\sum\limits_i p^0_i=\sqrt{s_0}$, where ${\mbox{\boldmath$p$}}_i$
denotes the three momentum of $i$-th produced particle.  In order to put recombined hadrons
onshell, we multiply the three momenta of all produced hadrons by a common factor $f$ 
so that the summation $ \sum\limits_i \sqrt{(f {\mbox{\boldmath$ p$}}_{i})^2 
+ m_{i}^2}$ would be equal to $ \sqrt{s_0}$\cite{tnnik}.

We take into account pseudoscalar $(Ps)$, vector $(V)$ and tensor $(T)$ mesons. 
The production probabilities for them are assumed to be $ P_{PS}=1/9,  P_V =3/9$ and $ P_T=5/9 $.
For baryons we consider lower lying baryons: 
 octet ($B_8$) and decuplet ($B_{10}$) baryons composed of $ u,d, $ and $ s $ flavors and 
the corresponding ones with charm flavor.   Octet and decuplet baryons are described as 
\begin{eqnarray}
|B_8 > = \cos\theta|q[q'q'']> + \sin\theta|q\{q'q''\}>,
\label{eqn:octB} 
\end{eqnarray}
\begin{eqnarray}
|B_{10} > = |q\{q'q''\}>.
\label{eqn:dcpB} 
\end{eqnarray}
We assume the production probabilities for octet and decuplet baryons from symmetric diquarks to be 
 $P_{B8}=1/3 $ and $ P_{B10}=1-P_{B8}=2/3$, respectively. 
Then the emission probabilities 
of individual cascade processes are determined from the above 
probabilities.  For examples, the probabilities to produce 
$~~\pi^+, ~\rho^+,~a_2^+,...,~\Delta^{++}, ~p,
\Delta^+,..., \Xi_{cc}^{++}$ and $\Xi_{cc}^{*++}~$ 
from a $~u~$ quark are as follows: \\

\noindent
$(1~-~\epsilon)~P_{d\bar{d}}~P_{PS},~~
(1~-~\epsilon)~P_{d\bar{d}}~P_V,~~
(1~-~\epsilon)~P_{d\bar{d}}~P_T,..., \\
\epsilon~ P_{u\bar{u}}^{~~2} ,~~ 
\epsilon~ P_{u\bar{u}} ~P_{d\bar{d}}~\frac{1}{3} ~P_{B8}~\sin^2\theta / (~\frac{1}{3}~ P_{B8} ~\sin^2\theta ~+ ~\frac{2}{3}~P_{B10})~,~~\\
\epsilon ~P_{u\bar{u}} ~P_{d\bar{d}}~ \frac{2}{3}~P_{B10} /~(\frac{1}{3} ~P_{B8} ~\sin^2 \theta ~+ ~\frac{2}{3} ~P_{B10} ),...,\nonumber  \\ 
\epsilon ~P_{c\bar{c}}^{~~2}~\frac{2}{3}~P_{B8} ~\sin^2 \theta/~(\frac{2}{3} ~P_{B8} ~\sin^2 \theta ~+ ~\frac{1}{3} ~P_{B10}~), \\
\epsilon ~P_{c\bar{c}}^{~~2}~\frac{1}{3}~P_{B10} /~(\frac{2}{3} ~P_{B8} ~\sin^2
\theta~+ ~\frac{1}{3} ~P_{B10}),\nonumber \\ 
$

\noindent
where the factors 1/3 and 1/2 are flavor SU(4) factors.
Directly produced resonances decay into stable particles.

The mixing angle of symmetric and anti-symmetric diquarks in octet baryon $\cos \theta$ and 
the pair creation probabilities of $u\bar{u}, d\bar{d}$ and $s\bar{s} $ 
are determined from the analyses of non-charmed 
hadron productions in $\pi p,~Kp$ and $pp$ collisions. In Ref.\cite{qdq}, 
we used $\cos^2\theta=\frac{2}{3}$ and  
$ P_{u\bar{u}}=0.45,~ P_{d\bar{d}}=0.45$ and $ P_{s\bar{s}}=0.10$. 
To reproduce the data on $\pi^-p \rightarrow D^\pm X$ cross section, we choose 
$\cos^2\theta=\frac{1}{2}, P_{c\bar{c}}=0.0003,~ P_{u\bar{u}}=0.435,~ P_{d\bar{d}}=0.435$ and
$ P_{s\bar{s}}=0.1297$\cite{tnkn}.
The dynamical parameters $\gamma_{[~]}$ and 
$\gamma_{\{\}}$ as well as 
the mixing angle of symmetric and anti-symmetric diquarks in octet baryon is related to 
the $\pi^\pm$ spectra in proton fragmentation region and are chosen as 
$\gamma_{[~]}=1.5$ and $\gamma_{\{\}}=2.0$\cite{tnnik,tnkn}. 

\subsection{Interaction with nucleus}

We now consider hadron-nucleus interaction. We regard the hadron-nucleus collision as a sum 
of hadron-nucleon collisions 
and neglect the intra-nuclear cascade of produced hadrons. The projectile hadron successively 
collides with nucleons inside the nucleus. The probability $P_{hA}(\nu)$ that the incident hadron $h$ 
collides with $\nu$ nucleons is calculated by the Glauber-type multiple collision model\cite{qdqA}. 
The number $\nu$ is determined by the distribution of nucleons in the nucleus and the cross 
section of the incident hadron with a nucleon $\sigma_{hN}$. 
We take the values $\sigma_{pN}=40, \sigma_{\Sigma^-N}=34, \sigma_{\pi^-N}=24$ and
$\sigma_{K^-N}=21$mb in the hadron-nucleus collisions. The nucleon number density of the 
nucleus with a mass number $A$ is assumed as
\begin{eqnarray}
\rho(r)=\rho_0\frac{A}{1+\exp(\frac{r-r_A}{d})},
\label{eqn:Wood-Saxon} 
\end{eqnarray}
where we choose $r_A=1.19A^\frac{1}{3}-1.61A^{-\frac{1}{3}}$ (fm) and  $d=0.54$ (fm).
\begin{figure}
\center
\includegraphics[height=13cm]{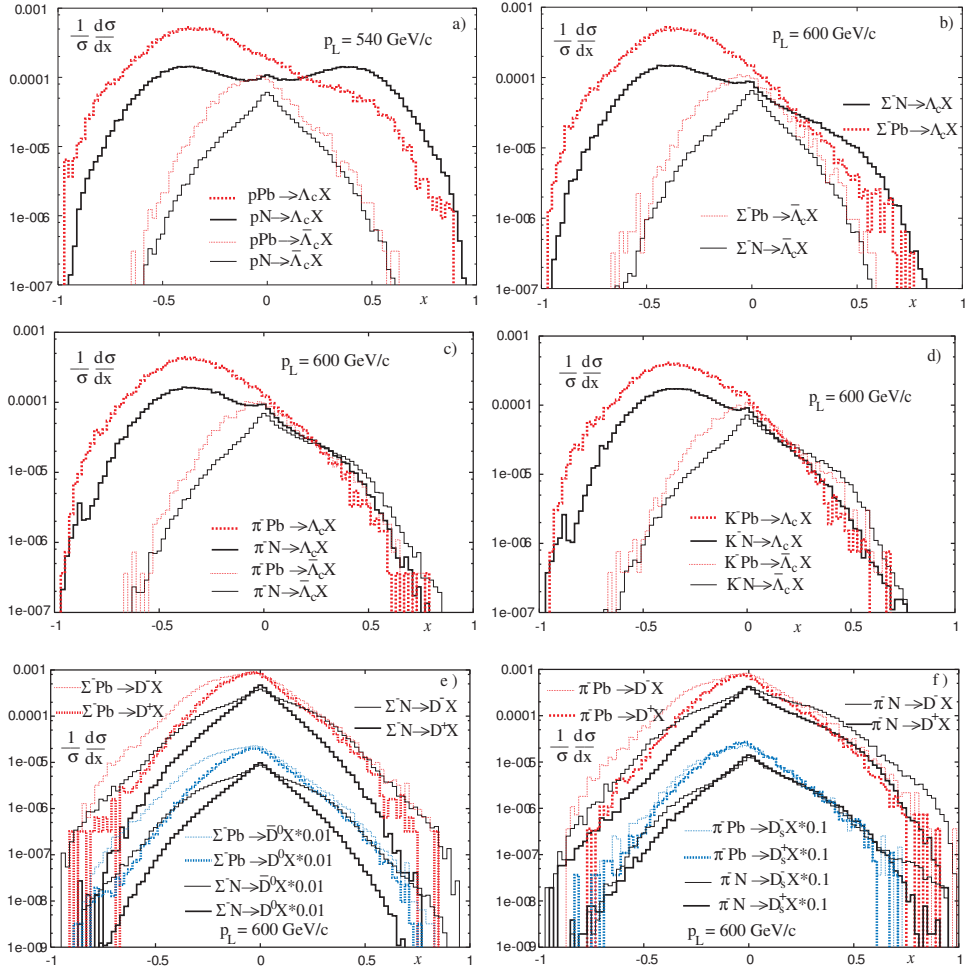}
   \caption{Feynman $ x $ distributions of $ \Lambda_c^+ $ and $ \bar{\Lambda}_c^- $ in a) $p$, b) $\Sigma^-$, 
c) $\pi^-$, d) $K^-$ and charmed mesons distributions in e) $ \Sigma^- $ and f) $\pi^-$ collisions with $N$ and $Pb$ targets. 
}
\label{fig:Lamc+-NPb}       
\end{figure}
Then it is assumed that $M_1N_1, M_2N_2,...,$ $M_{\nu-1}N_{\nu-1}$ and $hN_\nu$ 
interactions take place, where $M_i$ is $q\bar{q}$ state emitted from $h$ as $M_A$ discussed 
in section \ref{sec:quark_dstb}. At each collision, the projectile hadron $h$ looses its 
momentum. From the data on $A$ dependences of the spectra of the projectile hadrons, we put 
the degradation rate of the momentum of $h$ as 
$P(z)=0.25z^{0.25-1}$ which is used in Ref.\cite{qdqA}.

\section{Comparison with experiment}

In this section, we give the results of our model for 
charmed hadron productions in $p, \Sigma^-, \pi^-$ and $K^-$ beam interactions with nuclei. 
To see the target mass number $(A)$ dependence, we show
the $x$ distributions of $\Lambda_c, \bar{\Lambda}_c, D$ and $\bar{D}$ productions in $p, \Sigma^-, \pi^-$ and $K^-$ 
interactions with 
$N$ and $Pb$ targets at $p_L=600$ GeV/c in Fig.2.  
All spectra increase with 
$A$ in the region $x \stackrel{<}{\sim} 0$.  
The spectra of $\Lambda_c$ 
become soft with $A$ in the baryon beam fragmentation region.
This is due to the momentum degradation of the incident baryon through the multiple interactions 
with nucleons in the target. 

\begin{figure}
\centerline{}
\includegraphics[height=8cm]{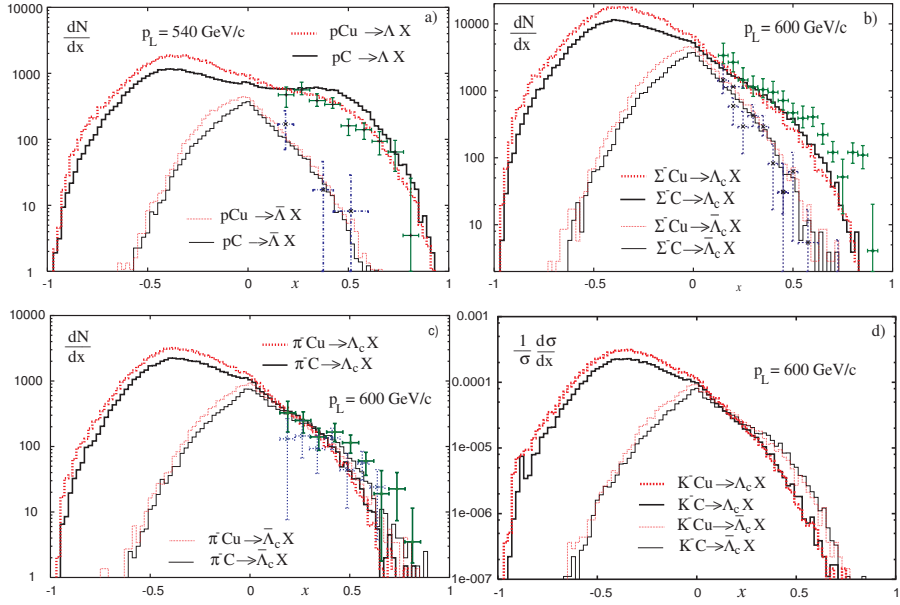}
   \caption{Feynman $ x $ distributions of $ \Lambda_c^+ $ and $ \bar{\Lambda}_c^- $ in a) $p$, b) $\Sigma^-$
, c) $\pi^-$ and d) $ K^- $ collisions with $C$ and $Cu$ targets in arbitrary unit.  The experimental data are taken from Ref.~21.
}
\label{fig:Lamc+-NPb}       
\end{figure}
The spectra of $\bar{\Lambda}_c$ 
scarcely depend on $A$ in the baryon beam fragmentation regions.  
In meson beam fragmentation regions, both $\Lambda_c$ and $\bar{\Lambda}_c$ slightly depend on $A$ 
and become soft with increasing $A$.
In baryon beam fragmentation regions, $A$-dependences of
the spectra of $\bar{D}$ mesons are smaller than those for $D$ mesons. The shapes of the spectra for $D$ mesons 
in baryon beam fragmentation regions are steep as compared with those of $\bar{D}$ mesons in baryon and $D$ and 
$\bar{D}$ mesons in meson beam fragmentation regions.

In Fig.3, we show the 
$x$ distributions of $\Lambda_c$ and $\bar{\Lambda}_c$ in $p, \Sigma^-, \pi^-$ and $K^-$ beam 
interactions with $C$ and $Cu$ targets in arbitrary unit 
and compare the results with the experimental data E781\cite{SELEX}. 
The shapes of $\Lambda_c$ and $\bar{\Lambda}_c$ in $p, \Sigma^-$ and $\pi^-$ fragmentation 
regions are well explained by our model.  
In our model, the shapes of the spectra are strongly related to the 
distributions of valence constituents in the incident hadrons, which are characterized by the 
parameters $\beta$'s in (\ref{eqn:Ha/A}).  
 Diquarks in the incident baryons have a tendency to bring a large momentum fraction of the incident 
baryons. The valence constituents $d$ and $\bar{u}$ in $\pi^-$ beam have the same distribution i.e. energetic 
$d$ and wee $\bar{u}$ or vice versa. The valence constituents $d$ and $\bar{u}$ in $\pi^-$ are 
harder than the valence quarks in $\Sigma^-$ and $p$ beams. The $ds$ diquark in $\Sigma^-$ is 
harder than the $ud$ diquark in $p$.  The valence $\{ud\}$ diquark in $p$ beam is the hardest among 
the commonly shared constituents between $\Lambda_c$ and the incident  $p, \Sigma^-, \pi^-$ and $K^-$ 
beams. 
Therefore $x$-distribution of $\Lambda_c$ in proton beam is harder than those in other hadron beams. 
\begin{figure}
\center
\includegraphics[height=8cm]{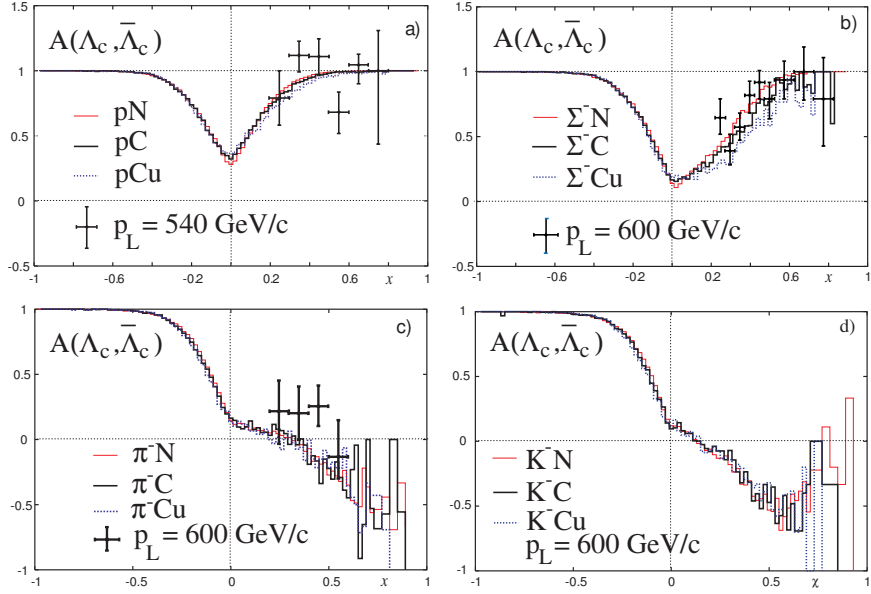}
   \caption{Asymmetries between $ \Lambda_c^+ $ and $ \bar{\Lambda}_c^- $ in a) $p$ 
b) $\Sigma^-$, c) $\pi^-$ and d) $ K^- $ collisions with $C$ and $Cu$ targets.  The experimental data are taken from Ref.~21.
}
\label{fig:Lamc+-NPb}       
\end{figure}

\begin{figure}
\center
\includegraphics[height=8cm]{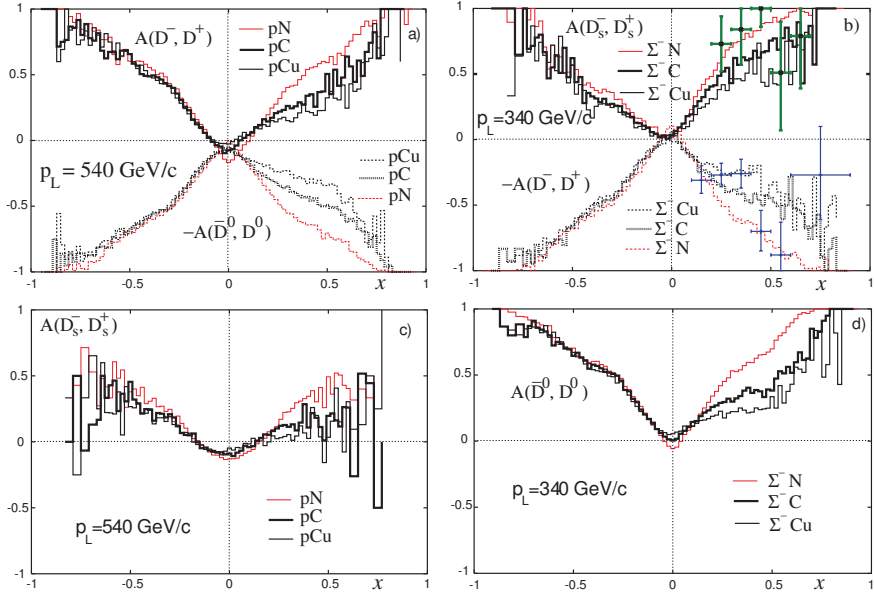}
   \caption{Asymmetries $A(D,\bar{D})$ between leading and non-leading charmed mesons in a) $pA$ and 
b) $\Sigma^-A$ collisions. Asymmetries between non-leading charmed mesons in c) $pA$ and 
d) $\Sigma^-A$ collisions.  The experimental data are taken from Ref.~20.
}
\label{fig:A_pSigma}       
\end{figure}

The asymmetries between the spectra $\Lambda_c$ and $\bar{\Lambda}_c$ in 
$pA, \Sigma^-A, \pi^-A$ and $K^-A$ collisions are shown in Fig.4. 
The results are in good agreement with the 
experimental data except for $\pi^-$ beam. 
For $\pi^-$ beam, $\Lambda_c$ and $\bar{\Lambda}_c$ share the common valence constituents 
$d$ and $\bar{u}$, respectively. Therefore the same spectra are 
expected for $\Lambda_c$ and $\bar{\Lambda}_c$ productions in the region of $\pi^-$ 
beam fragmentation. 
In our model, however, the excess of $\bar{\Lambda}_c$ over $\Lambda_c$ 
is seen at $0.3 \stackrel{<}{\sim} x$.  This is due to the fact that energetic $\Lambda_c$ and $\bar{\Lambda}_c$ are 
produced in the $d$-quark-diquark and the $\bar{u}$-antiquark-quark chains, respectively. 
The former chain is slightly shifted to the target fragmentation region as compared with the 
latter one due to the large momentum fraction of the diquark\cite{tnnik}.
Therefore our model gives the negative asymmetry $A_{\pi^-A}(\Lambda_c,\bar{\Lambda}_c)$ at $0.3 \stackrel{<}{\sim} x$.
The valence diquark in $\Sigma^-$ beam contributes to $\Lambda_c$ production via a processes such as 
$ds \rightarrow du + K^- \rightarrow \bar{c} + \Lambda_c + K^-$ in $\Sigma^-$ fragmentation region. 
The contributions to $\Lambda_c$ production from the valence diquarks dominate the contributions 
from the valence quarks in $\Sigma^-$ beam.

In Fig.5, We show the results of asymmetries between charmed and anti-charmed 
mesons in $p$ and $\Sigma^-$ interactions with $N, C$ and $Cu$ targets.
The asymmetries $A_{pA}(D^-,D^+), A_{pA}(\bar{D}^0,D^0), 
A_{\Sigma^- A}(D_s^-,D_s^+)$ and $A_{\Sigma^- A}(D^-,D^+)$ are 
small as compared with the asymmetries $A_{pA}(\Lambda_c,\bar{\Lambda}_c)$ and $A_{\Sigma^- A}(\Lambda_c,\bar{\Lambda}_c)$.  
Although both $D_s^+$ and $D_s^-$ in $p$ and $ D^0$ and $\bar{D}^0$ in $\Sigma^-$ beam are 
non-leading particles, the results of asymmetries $A_{pN}(D_s^-,D_s^+)$ and 
$ A_{\Sigma^- A}(\bar{D}^0,D^0)$ are 
non-zero and have large values. This is because in our model $\bar{D}$ mesons are produced 
mainly from the chain between the valence diquark in the incident baryon and the valence quark 
in the target nucleon. For example, $D_s^-$ in proton fragmentation region is produced via a process 
such as $ud \rightarrow sd + K^+ \rightarrow cd + D_s^- + K^+$.
The more energetic the valence diquark becomes, the more copiously $\bar{D}$ mesons are produced 
in the baryon fragmentation region. 
In our model the valence $ds$ diquark in $\Sigma^-$ beam 
has a larger momentum fraction than the valence diquarks in $p$ beam. 
Consequently the asymmetries $A_{\Sigma^-A}(\bar{D}^0,D^0)$ are large as compared with $A_{pA}(D_s^-,D_s^+)$. 
The asymmetry $A_{\Sigma^- A}(D_s^-,D_s^+)$ is larger than other
charmed meson asymmetries in $p$ and $\Sigma^-$ beams. This is also due to the large momentum fraction 
of the valence $ds$ diquark in $\Sigma^-$. In the target fragmentation region,
both $D$ and $\bar{D}$ mesons increase similarly with 
the mass number $A$ and the asymmetries show small mass number dependences.
Since the spectra of $D$ mesons increase with $A$ and those of 
$\bar{D}$ mesons change little, 
there are considerable target mass number dependence in the asymmetries $A(\bar{D},D)$ 
at $0 \stackrel{<}{\sim} x$.

\section{Conclusions}
\label{sec:4}

  1. We have examined hadroproductions of charmed baryons in the frame work of the 
quark-diquark cascade model. Their spectra are 
well explained by using the values of the dynamical parameters fixed from Regge intercepts.
The large leading/non-leading asymmetries $A_{pA}(\Lambda_c,\bar{\Lambda}_c)$ and 
$A_{\Sigma^-A}(\Lambda_c,\bar{\Lambda}_c)$ are naturally explained. 

2. The valence diquark in the incident baryon, 
which brings a large momentum fraction of the beam, plays an important role for $\bar{D}$ 
productions as well as for $\Lambda_c$ production\cite{qdq}. 
 The non-leading anti-charmed meson productions in baryon beams are largely 
affected by the valence diquark distribution in the incident baryons through the processes 
(13) and/or (15).

3. In meson beam fragmentation regions, 
$A$-dependences of the spectra of non-leading charmed mesons are small as compared with 
those for leading charmed mesons.
There are considerable $A$-dependences in $\Lambda_c$ and $D$ meson 
productions, while those for $\bar{\Lambda}_c$ and $\bar{D}$ 
productions are small in baryon beam fragmentation regions.

4. There are discrepancies between the data and calculations of asymmetries 
$A_{\pi^-A}(\Lambda_c,\bar{\Lambda}_c)$ at $\pi^-$ fragmentation region as seen above. 
Piskounova discussed this problem successfully by introducing the transfer of string 
junction in quark-gluon string model\cite{piskounova_Lambda_c}.
To explain the experimental data in our model, it is necessary, for example, a energetic 
$d$ quark from $\pi^-$ 
beam and a wee $uc$ diquark from target nucleon to recombine into a $\Lambda_c$. 
Here the $uc$ diquark is converted from the valence diquark in the target nucleon by 
emitting several mesons through the processes (13) and/or (15).
The significant value of the asymmetry 
$A_{\pi^-A}(\Lambda_c,\bar{\Lambda}_c)$ as well as the small value of 
$ A_{\pi^- N}(D^0,\bar{D}^0)$ in meson beam fragmentation 
is a challenging problem.

\section*{Acknowledgments}

This work was partly supported by grant of the Academic Frontier Project 
promoted by the Ministry of Education, Culture, Sports, Science and Technology, Japan.






\end{document}